\documentclass[onecolumn]{aastex631}

\newcommand{\lsst}{\textit{LSST}}
\newcommand{\gaia}{\textit{Gaia}}
\newcommand{\AU}{\mathrm{au}}

\newcommand{\eqq}[1]{Equation~(\ref{#1})}
\newcommand{\ie}{\textit{i.e.\/}}
\newcommand{\eg}{\textit{e.g.\/}}
\newcommand{\vecb}{\mathbf{b}}
\newcommand{\vecq}{\mathbf{q}}
\newcommand{\vecx}{\mathbf{x}}
\newcommand{\vecv}{\mathbf{v}}
\newcommand{\vecr}{\mathbf{r}}
\newcommand{\matF}{\mathsf{F}}

\usepackage{natbib} 
\usepackage{amsmath}
\usepackage{enumitem}
\usepackage{verbatim}
\usepackage{graphicx}
\usepackage{subfigure}
\usepackage{color}
\usepackage{xcolor}
\usepackage{float}
\usepackage{hyperref}

\shorttitle{Asteroid masses from LSST}

\begin{document}

\title{Determinations of asteroid masses using mutual encounters observed in the Legacy Survey of Space and Time}

\author[0000-0002-8613-8259]{Gary M. Bernstein}
\affiliation{Department of Physics and Astronomy, University of Pennsylvania, Philadelphia, PA 19104, USA}
\email{garyb@physics.upenn.edu}
\correspondingauthor{Gary M. Bernstein}

\author[0009-0006-4072-6385]{Negin Najafi}
\affiliation{Department of Physics and Astronomy, University of Pennsylvania, Philadelphia, PA 19104, USA}
\email{neginn@upenn.edu}

\author[0000-0001-6299-2445]{Daniel C. H. Gomes}
\email{dchgomes@gmail.com}
\affiliation{Department of Physics and Astronomy, University of Pennsylvania, Philadelphia, PA 19104, USA}

\begin{abstract}
  Using published simulations of the 10-year \textit{Legacy Survey of Space and Time} (\lsst), we forecast its ability to determine the masses of individual main-belt asteroids (MBAs) through precise astrometry of any pairs of the $\approx1.2$~million known MBAs undergoing close gravitational encounters during the survey.  The uncertainty $\sigma_I$ on the impulse applied to a tracer asteroid by its deflector is derived from the Fisher matrix of the tracer's astrometric data, including an azimuthal acceleration $A_2$ from the Yarkovsky effect as a free parameter for each tracer.  If only \lsst\ observations are available, $\sigma_I \approx7\times10^{-6}\,{\rm m}\,{\rm s}^{-1}$ for MBAs at apparent magnitude $m_V<19.5,$ degrading $\approx10\times$ for $m_V=23.$ These tracers yield a median uncertainty on the mass of an MBA of $\approx4\times10^{-14}M_\odot,$ with a wide range of variation depending on the ``luck'' of close encounters.  Roughly 125 MBAs obtain mass measures with $S/N>5.$  If pre-\lsst\ astrometry yields a strong constraint on the state vector of the tracer MBA at the start of \lsst, then these values improve to median $\sigma_M\approx1.3\times10^{-14}M_\odot$ and 310 MBAs at $S/N>5,$ with $>1/2$ of these having $S/N>10.$  These yields would be a $\approx10$-fold increase in the number of known asteroid masses, including a nearly complete knowledge of MBAs with $H<7.5.$  If pre-\lsst\ data are sufficient to start constraining the Yarkovsky effect, another factor $\sim1.5$ can be gained.  Tables of the measurable deflector MBAs and their tracers are provided.
\end{abstract}


\section{Introduction}
Determination of the masses of asteroids is essential to understanding their composition and history, and for a complete understanding of the dynamics and precision ephemerides of the Solar System.  There are multiple methods to measure a minor planet's mass without assuming a density: high-precision values can usually be inferred for bodies having well-tracked natural satellites or visiting spacecraft.  As summarized by \citet{inpop}, spacecraft data have yielded masses for just 5 main-belt asteroids (MBAs): (1) Ceres and (4) Vesta, whose summed mass of $1.197\times10^{21}$~kg is roughly half of the total main-belt mass; plus (21) Lutetia, (243) Ida, and (253) Mathilde.  Natural satellites' orbits have given masses for another 12 MBAs.
% Kalliope, Daphne, Eugenia, Sylvia, Antiope, Camilla, Hermione, Elektra, Kelopatra, Emma, Alauda, Pulcova
The satellite-based determinations generally yield masses at 1\% accuracy or much better.  A second method is to infer asteroid masses from an ephemeris solution to high-precision astrometry and ranging of the major planets, with the cm-accuracy ranging to Mars being of the most power.  \citet{inpop} use this method to detect masses at $S/N>3$ for 104 main-belt asteroids.  They obtain sub-percent precision ($S/N>100$) on Ceres and Vesta, and $<10\%$ uncertainty ($S/N>10$) on 7 other bodies (including Pallas and Hygeia, the next-most-massive bodies).  The JPL DE440 ephemeris \citep{de440} and the EM2017 ephemeris \citep{pitjeva} create precision ephemerides including $\approx300$ MBAs as gravitating bodies, but the masses of these bodies are either held fixed to values determined by other means, or do not have uncertainties reported.

The third method for determining asteroid masses, and the one we focus on in this paper, is by fitting to precision astrometric data for pairs of asteroids that undergo a mutual close encounter.  Adopting the impulse approximation for these encounters, if the astrometric data can determine the size $I$ of the impulse applied to the \emph{tracer} body, then we can infer the mass of the \emph{deflector} body with knowledge of the relative speed $v$ and the separation $\vecb$ of the bodies at the time of closest approach.  Approximating the bodies' motions as inertial through the encounter yields the standard formula
\begin{equation}
  I = \frac{2GM}{bv}.
  \label{impulse}
\end{equation}
The mutual-encounter method has been used successfully for roughly 50 years \citep{Hertz}.  \citet{baer2011} report masses at significance $S/N\ge3$ for 26 MBAs derived from 82 encounters.  Of these, 3 obtain sub-percent accuracy (Ceres, Vesta, and (15) Eunomia), and 9 more yield $S/N>10.$  \citet{Zielenbach} fits for masses of 104 MBAs, using a large number of tracer bodies for each, obtaining $S/N>100$ for Ceres and Vesta, $10<S/N<100$ for 9 MBAs, and $3<S/N<10$ for 22 more asteroids.

The review by \citet{hilton} reports just 24 known asteroid masses at that time, with some increase in the following decade from the methods described above.
Advances in observational techniques in the 2010's and 2020's should, however, yield a substantial expansion in the number and accuracy of mass determinations from mutual encounters.  Firstly, systematic sky surveys have vastly increased the number of known asteroids, \ie\ potential tracer particles, to over 1 million.  Secondly, the \gaia\ Data Release 3 \citep{gaiass3} included mas-precision positions for $>23$~million epochs of $>150,000$ solar system bodies with apparent magnitudes $G<21$ over 34 months of operation, extended to 66 months in \citet{gaiafprsso}.  While the number of new asteroid masses reported from \gaia\ DR3 has been modest \citep{Zach,liDR3} to date because of the short time duration of released data, the \gaia\ spacecraft operated for $>10$ years, and later data releases should yield a larger number of mass detections as well as better mass accuracy per event.  Perhaps more importantly, Gaia has defined a stellar reference frame that allows ground-based observatories to place asteroid positions onto an inertial coordinate system with systematic errors well below 1~mas.  The \textit{Legacy Survey of Space and Time (LSST)} to be conducted with the \textit{Vera Rubin Observatory} from 2025--2035 will detect and track millions of MBAs, with accuracy limited by a combination of (magnitude-dependent) photon noise and $\approx 2$~mas uncertainty from refraction by atmospheric turbulence \citep{willow}.  Each source will be detected 200--600 astrometrically useful times over the course of the \lsst.   The per-observation astrometric uncertainties for \lsst\ are smaller than those for \gaia\ at the faint limit of the latter, and \lsst's observations will extend $\approx 3$ mag deeper, increasing the pool of potential tracers by $\approx100\times.$ As noted in the \textit{LSST Science Book} \citep[][Section~5.4.3]{lsstbook} ``At present only a few dozen asteroids have mass estimates based on perturbations, but \lsst\
will produce astrometry that is both prolific and precise, at the same time that it dramatically
expands the pool of potential test particles. \lsst\ data should allow the estimation of the mass of
several hundred or so main belt asteroids with an uncertainty of $\sim30\%$ or less.''

In this paper, with \lsst\ now on the verge of operation, we quantify this prediction more carefully by finding all encountering pairs among all MBA and Jupiter Trojan asteroids (henceforth simply ``Trojans'') that should yield deflections of $\approx 1$~mas or greater over the 10-year observing period of \lsst, and using a simulation of \lsst's observing schedule and performance to forecast the uncertainties in the masses of the deflectors in these encounters.  While the exact sequence of observations and astrometric errors of \lsst\ are of course not yet known, we are at this time able to make reasonable estimates of the future performance of the survey for this purpose, to know what fractions of the asteroids of what masses will be measured to useful significance using presently-known bodies as tracers.  For purposes of inferring an individual asteroid's composition via its density, we will consider a $1\sigma$ error of $<20\%$ on the mass, or $S/N>5,$ as the point where the mass information starts to become ``useful.''  One we reach $S/N\approx 10,$ additional precision on the mass will not generally lead to greater compositional knowledge, since uncertainties in the albedo/diameter/volume of the body will dominate the density uncertainty.  For dynamical purposes, however, it is the typical uncertainty $\sigma_M$ of the MBA's masses (rather than the $S/N$) that determines errors in ephemerides.

The forecast for knowledge of asteroid masses arising from mutual encounters during \lsst's 10-year lifetime still depends fairly strongly on the quality of pre-\lsst\ observations of the tracer body in each event.  At the pessimistic ``\lsst-only'' limit, the pre-\lsst\ data offer no useful information, and \lsst\ astrometry must solve for the state vector $(\vecx_0,\vecv_0)$ components at the start of \lsst\ and the transverse Yarkovsky acceleration coefficient $A_2$ for each tracer, reducing the accuracy on the impulse $I$.  In the optimistic limit, pre-\lsst\ data are strong enough to consider $\vecx_0, \vecv_0,$ and $A_2$ of the tracer to be known exactly, and \lsst\ is used only to measure the perturbation from the impulse.  We will consider an intermediate case where the initial state vector is known, but the Yarkovsky acceleration is not---we suspect this middle case is closest to reality for most MBAs.

We are not claiming to be exhaustive in our inventory of bodies whose masses \lsst\ will determine, since the precise list will depend upon the exact realization of the survey, and the results may also be influenced by tracer asteroids that are not yet discovered.  We also have not attempted to evaluate the accuracy of pre-\lsst\ observations for each tracer body, but it is clear that this will be a critical activity for getting the most asteroid-mass knowledge out of \lsst.

\section{Identifying encounters}
The search for asteroid-asteroid encounters with potentially detectable effects on \lsst\ astrometry begins with the \texttt{MPCORB.DAT} file containing $H$ values and osculating heliocentric orbital elements for all asteroids known to the Minor Planet Center (MPC).\footnote{Downloaded 6 Jan 2025 from the \href{https://www.minorplanetcenter.net/data}{MPC data page}.}  We retain from this file the 1.26~million objects with multiple oppositions of observations, semi-major axes $a<6\,\AU$ and MPC uncertainty indicator $U\le 5.$  We also exclude Ceres, Vesta, and Pallas from the list since they will generate very large numbers of encounters above our impulse threshold that are not of interest---the first two already have their masses very well determined by spacecraft, and Pallas is already approaching percent-level accuracy from previous data on mutual encounters.

The osculating heliocentric orbital elements are then converted to barycentric state vectors in the ICRS coordinate frame, and integrated forward in time to the start date of our simulated \lsst\ survey using a simple leapfrog integrator with time step of $\Delta t=1$~day.  The integrator assumes Newtonian dynamics, with accelerations derived from the positions of the Sun and 8 planets.  Masses and positions for the gravitating bodies are taken from the JPL DE440 ephemeris.\footnote{Downloaded from \href{https://naif.jpl.nasa.gov/pub/naif/generic\_kernels/spk/planets/}{this JPL site}.} While \emph{measurement} of asteroid masses from real data with mas accuracy will require more sophisticated integrators, incorporating relativistic effects and perturbations from some of the larger asteroids, the \emph{forecasting} of inferences does not require great accuracy.  The ephemeris accuracy must only be sufficient to successfully predict (1) the impact parameters and relative velocities of encounters to $\lesssim10\%$ accuracy, and (b) the \emph{derivatives} of the observed position with respect to initial state vector and the impulse velocity, to similar accuracy.  Because the closest encounters have $b\approx10^{-5}\,\AU,$ integrator accuracy of $10^{-6}\,\AU$ or $\approx150$~km is more than adequate.

We wish to find all encounters during the 10-year survey that will generate a statistically significant shift in the tracer's \lsst\ measurements.
Based on estimated accuracy of $\gtrsim1$~mas for \lsst\ astrometry of tracers, we make a rough estimate that encounters generating impulses of $I < I_{\rm min}=2.2\times10^{-6}\,{\textrm m}\,{\textrm s}^{-1}$ will not be detectable at $\gtrsim3\sigma$ in the astrometry of a tracer. 
From \eqq{impulse} we can derive a maximum impact parameter $b_{\rm max}$ between a deflector of mass $M$ and a tracer moving at relative velocity $v_{\rm min}$ that will result in an impulse above $I_{\rm min}.$  Initial investigation of the distribution of relative velocities in encounters shows that
setting $v_{\rm min}=100 \,{\textrm m}\,{\textrm s}^{-1}$ will exclude only a handful of encounters over a decade---and in these cases the encounter time would no longer be a small fraction of the orbital period anyway, so these would need to be studied individually as coupled bodies.  We obtain
\begin{equation}
  b_{\rm max}(H) = \frac{2GM}{I_{\rm min} v_{\rm min}} = 0.09\times 10^{-0.6(H-10)}\,\AU,
  \label{bmax}
\end{equation}
We have assumed above that the asteroid's $H$ determines its mass, under a model of spherical bodies with uniform albedo (0.25) and density (2500~kg~m$^{-3}$), such that $H=10$ corresponds to $2.3\times10^{16}\,\textrm{kg}\, \approx 10^{-14}M_\odot.$ For asteroids that are denser and/or darker than these nominal values, we may miss some encounters at $b>b_{\rm max}$ that still yield detectable tracer displacements.  We will limit $b_{\rm max}(H)$ to a value of $0.1\,\AU$ for $H<10$ deflectors---our results suggest that it is rare for the most informative encounter for a given deflector to be an event with $b>0.1\,\AU.$

The asteroids are split into groups spanning intervals of 1~mag in $H$, and $b_{\rm max}(H)$ is evaluated at the brightest $H$ value for each group.  
We next enter a loop to identify encounters occurring during the time period 1 May 2025 through 1 May 2035 of the simulated \lsst.\footnote{The real \lsst\ is currently scheduled to begin later in 2025; this will change the details of individual asteroids' mass measures, but not the statistics of the full population.}
The loop begins with $t$ at the start date of the simulated survey.  Each iteration contains these steps:
\begin{enumerate}
\item If $t$ is after the survey end date, end the loop.
\item Build a $k$-d tree of the spatial positions $x(t)$ of the asteroids in each bin of $H$.
\item Use the (fast) $k$-d tree algorithms for finding close pairs between each combination of deflector $H$ bin and tracer $H$ bin.  The matching radius is set to
  \begin{equation}
    r_{\rm max} = \sqrt{ {\rm min}[0.1\,\AU, b_{\rm max}(H_{\rm deflector})]^2 + (v_{\rm max} t_{\rm step}/2)^2},
    \label{rmax}
  \end{equation}
  where $v_{\rm max}=3\times10^{4}\,\textrm{m}\,\textrm{s}^{-1}$ is the largest expected relative velocity for an interesting encounter, and $t_{\rm step}=2$~days is the time step for the loop.  This expression is the relative separation between deflector and target asteroid at time $t$, assuming that they reach closest approach at time $t_{\rm imp}$ such that $|t-t_{\rm imp}|\le t_{\rm step}/2.$  
\item For each (deflector,tracer) pair found within $r_{\rm max},$ we find the time $t_{\rm imp},$ and vector displacement $\vecb$ of closest approach, and relative velocity $v,$ assuming inertial motion with the positions $x(t)$ and velocities $v(t)$ of the two asteroids.
\item If $|t_{\rm imp}-t| > 0.55 t_{\rm step},$ \ie\ the closest approach is not within this time step, we discard the pair.  It should be found again and retained at an earlier or later time step.
\item We calculate the expected impulse $I$ generated by the encounter, using the mass derived from the deflector's $H$.  If $I < I_{\rm min},$ we discard the pair.
\item We use the leapfrog integrator to advance the state vectors and the time $t$ by $t_{\rm step}=2$~days.
\end{enumerate}

Only those encounters occurring between 1 and 9 years after survey start are retained, on the assumption that the others have insufficient data before or after the deflection to allow a useful differential measure.  In Section~\ref{results} we will calculate the actual uncertainty on $I$ for each encounter, marginalizing over the initial state vector.

\begin{figure}
  \plottwo{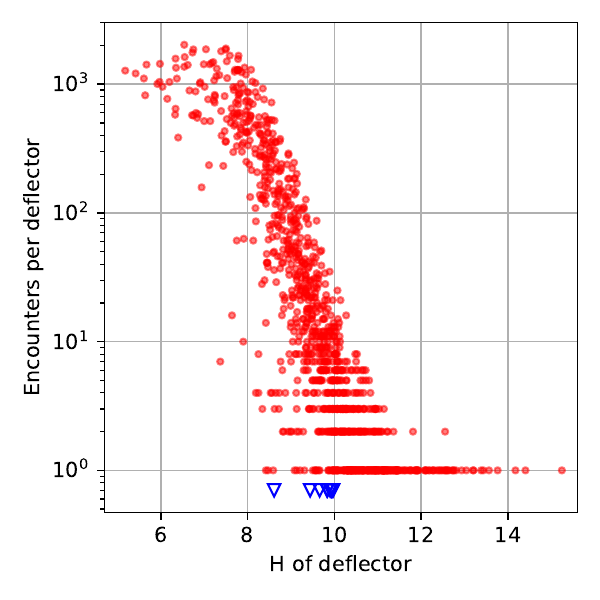}{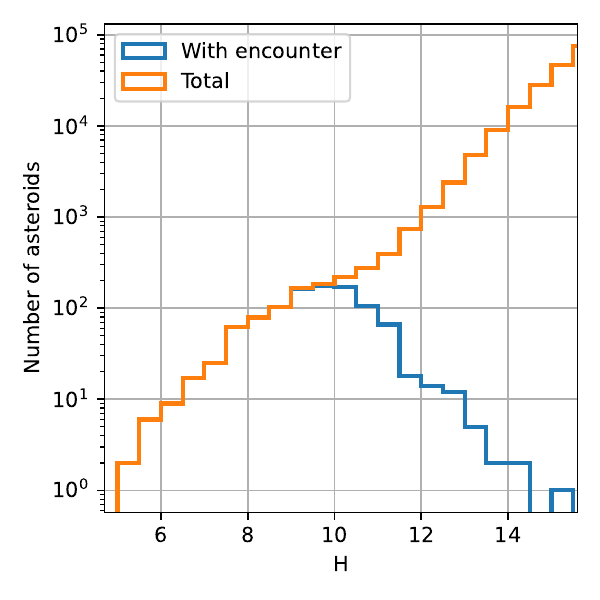}
  \caption[]{\small The results of the search for measurable deflector-target encounters in a simulated \lsst\ are plotted vs the absolute magnitude $H$ of the deflector.  The left-hand plot gives the number of identified encounters for each unique deflector---this number is well above unity for $H\lesssim 9.$  The 10 blue triangles at the bottom mark the $H$ values of 10 MBAs with $H<10$ that do not impart impulses above $I_{\rm min}$ on any known tracers.  The right-hand plot shows, in bins of 0.5~mag in $H$, the total number of known asteroids at $a<6\,\AU,$ and the number that have at least one encounter  perturbing a tracer with an impulse $>I_{\rm min}=2.3\times 10^{-6}$~m/s.    As expected from the left-hand plot, virtually all of the 725 MBAs at $H\lesssim10$ have at least one encounter above the impulse threshold during the \lsst, and another $\approx 400$ bodies at $10<H<15$ get ``lucky'' enough to encounter a target with small enough impact parameter to surpass the impulse threshold.}
  \label{encounters}
\end{figure}

The results of the search for encounters with impulse above $I_{\rm min}$ are illustrated in Figures~\ref{encounters}.  Only 10 of the 725 MBAs with $H<10$ fail to have potentially measurable encounters with a known target.  Indeed, most $H<10$ MBAs perturb many targets, into the thousands, even though we have truncated our search to $b<0.1\,\AU.$ In further analyses, we will only examine the 20 encounters with any given deflector that generate the highest impulses, since we expect that most of the measurable information on its mass will be provided by tracers receiving these top 20 impulses.

At $H\gtrsim11,$ a deflector starts to require some good fortune to encounter a target and deflect it measurably.  The cross-section for a measurable impact shrinks as $b_{\rm max}^2 \propto M^2 \propto 10^{-1.2H},$ which is steeper than the growth in $dN/dH\propto 10^{0.5H}$ in the number of the potential MBA deflectors, so the number of detectable encounters drops fairly quickly with deflector $H$ and mass.  Another $\approx400$ asteroids at $10<H\lesssim15$ generate impulses above $I_{\rm min},$ \ie\ down to masses of $\approx 1\times10^{-16} M_\odot.$ We will find however that the smallest asteroids attaining useful $S/N$ on the impulses and mass are are near $H=12.$ Among the Trojans, $\approx40$ generate above-threshold encounters, but none of them attain $S/N>5$ on their masses in any forecast, so we restrict our further discussion to MBAs.

If we now restrict our analysis to the 20 largest-impulse encounters for any deflector having $>20$ encounters on the candidate list, we are left with 6839 encounters involving 1081 unique deflectors.  There are 6800 unique tracers involved in these encounters---a small number (37, or 0.5\%) are involved in 2 (or more) encounters.  Ten of these are multiple encounters of the same deflector-target pair, which requires the deflector and target to be on very similar orbits.  Since 6800 of the 1.2 million possible tracers, or 0.5\%, are involved in 1 encounter during \lsst, we would expect 0.5\% of these 6800, or $\approx34,$ to have 2 encounters, if encounters are uncorrelated events (Poisson statistics).  This is consistent with what we find.  In futher analysis we will take the shortcut of analyzing each encounter of a given tracer independently, neglecting the covariances between mass estimates from these rare multiple-encounter cases.

\section{Dynamical model}
The next step is to quantify the observational signatures of deviations from the nominal MPC orbit for each of the 6839 encounters.  Free parameters of the tracer orbit are the six elements of the state vector at the start of \lsst, plus the applied impulse $I$.  We need also to consider non-gravitational forces on the target body if (1) the astrometric deviations they cause, or similarly their integrated impulse on the target, are larger than $\approx1/10$ of the effects the gravitational encounter, \emph{and} (2) the time signature of the effect resembles that of an impulse, such that a fit to the data would find the non-gravitational force covariant with the size of $I$.

Radiation pressure satistifies criterion (1) for smaller targets.  The acceleration $a_{\rm rad}$ from incident radiation pressure on a spherical body of radius $R$ and density $\rho$ at distance $r$ from the Sun is
\begin{eqnarray}
  a_{\rm rad} & = & \frac{\pi R^2 L_\odot}{4\pi r^2 c} \frac{3}{4\pi R^3 \rho} \nonumber \\
              & = & \frac{3 L_\odot}{16\pi R\rho c} \frac{1}{r^2} \label{eq:arad1} \\
              & \approx & 1.7\times10^{-12}\times 10^{0.2(H-16)}\,\text{m}\,\text{s}^{-2} \times \left(\frac{1\,\AU}{r}\right)^2,
\label{eq:arad2}                          
\end{eqnarray}
where the last line inserts our nominal values for asteroid albedo and density.  Applying this force over 10 years yields an impulse of $I_{\rm rad}\approx 5\times10^{-5}\,\text{m}\,\text{s}^{-1},$ which comparable to or larger than the typical uncertainties in gravitational impulses we will see in later sections.

The \emph{incident} radiation pressure does not, however, satisfy criterion (2), since a $1/r^2$ radial force does not have observable consequences similar those of an instantaneous impulse.  The orbit is altered only by being at very slightly larger distance for a given period, at a level that is undetectable in \lsst\ astrometry.  Similarly, the reflex pressure from \emph{reflected} sunlight should yield a nearly radial $1/r^2$ force and can be ignored here.  More intrusive, however, is the net reaction force from anisotropic thermal \emph{reradiation} of absorbed solar flux, \ie\ the Yarkovsky effect.  The component along the radius vector $\hat\vecr$ once again is not covariant with a gravitational impulse, but the component $a_Y$ directed in the azimuthal direction (perpendicular to $\hat\vecr,$ in the orbital plane) must be considered.

The component of an impulse that generates the largest observable signal is that parallel to the tracer's velocity (nominally azimuthal), which changes its energy, hence its semi-major axis $a$, and its period $P.$  This causes a perturbation in the tracer's ecliptic longitude that grows linearly with time after the impulse.  The azimuthal Yarkovsky force changes $a$ at a constant rate, leading to an azimuthal perturbation that grows quadratically with time.  This quadratic Yarkovsky signal is covariant with the zero-then-linear signal from an encounter, degrading the accuracy of the inference on $I$.  Longer periods of observation make it easier to disentangle these effects.

We therefore include an azimuthal Yarkovsky acceleration $a_Y=A_2(1\,\AU/r)^2$ in the dynamical model of the tracer.  We can in general write
\begin{equation}
  a_Y = \gamma g_Y a_{\rm rad},
  \label{eq:aY}
\end{equation}
where $\gamma$ is the fraction of incident radiation absorbed at the surface, and $g_Y$ is a factor giving the average azimuthal component of the angular radiation pattern.  For a spherical body with a homogeneous, Lambertian surface, the maximum attainable value of $g_Y$ is $\approx0.1,$ for the diurnal Yarkovsky effect \citep[\eg][Appendix H]{tremaine}. The maximum is attained when the spin axis of the asteroid has obliquity $\epsilon=0$ or $\pi,$ \ie\ perpendicular to the orbital plane, and the thermal/rotation properties reach an optimum value.  The diurnal effect scales as $\cos\epsilon,$ so can be either a prograde or a retrograde force, just as a gravitational impulse can be.  An isotropic distribution of spin axes would yield an RMS variation of $a_Y$ equal to $1/\sqrt{3}$ of its maximal value, and variations in thermal properties from the optimum would lower this further.

There are other subtleties to the Yarkovsky force---a seasonal effect, binary effects, and results of non-Lambertian surfaces \citep[see review by][]{yarkovsky}---but these would typically enter at lower amplitude and/or with temporal patterns less resembling those of an impulse.
We will therefore limit our Yarkovsky treatment to having a single free parameter $A_2$ for each tracer, and will place a prior probability distribution on $A_2$ that is Gaussian with a standard deviation implied by Equations (\ref{eq:arad2})  and (\ref{eq:aY}) with the $H$ value of the tracer and a value of $\gamma g_Y = 0.05.$  The resultant typical integrated impulses from Yarkovsky effect are therefore $20\times$ smaller than that of the full incident radiation pressure $I_{\rm rad}$ estimated above---but still have significant impact on forecasted mass measurements, as we shall quantify below.

We then conduct another leapfrog integration of each tracer asteroid's path during the time period of the simulated \lsst.  In this integration we consider the tracer subject to the Yarkovsky force in addition to the gravity of the Sun and 8 planets.  We want the partial derivatives of the tracer's right ascension and declination $(\alpha,\delta)$ with respect to the parameters  $\vecq=\{x_0,y_0,z_0, \dot x_0,\dot y_0,\dot z_0,  I, A_2\},$ as function of time $t$ of observation.  We obtain this by integrating 8 test particles that have had the nominal MPC orbit perturbed by small deviations to each parameter.  This yields a time-dependent $2\times8$ derivative matrix
$\partial [\alpha(t),\delta(t)] / \partial \vecq.$  The apparent $V$-band magnitude of the tracer asteroid is saved as well, for purposes of forecasting the observational errors. We ignore phase corrections for this work.

\section{Forecasting \lsst\ observations and impulse uncertainties}
With a list of interesting encounters in hand, we must now quantify the ability of \lsst\ to measure the positions of the relevant tracer asteroids. We use the table of simulated \lsst\ exposures given in \texttt{baseline\_v4.0\_10yrs.db}, produced by the Rubin Observatory Survey Cadence Optimization Committee group.\footnote{Simulation files are linked and documented at \href{https://survey-strategy.lsst.io/baseline/index.html}{the Committee's site}.}  The simulations emulate a sequence of all exposures taken for the main ``Wide Fast Deep'' survey, and the circumstances of each, including time, sky coordinates, filter, sky conditions, and PSF size.  The methods for finding the exposures containing a given asteroid, and estimating the RMS measurement error $\sigma_\theta$ on each axis of its sky position, are the same as used by \citet{trojans}.  In brief: for each tracer MBA, we find all \lsst\ exposures that would contain its image.  The astrometric uncertainty that would be attained on the MBA has two components.  The first is a photon-noise error, which can be calculated using the estimated quantum efficiency of the observatory along with the MBA's apparent magnitude, and the circumstances of the (simulated) exposure: filter band, sky background, cloud cover, airmass, and width of the point-spread function.  The second component is a fixed 2~mas, added in quadrature to the photon-noise error to account for \lsst\ astrometric calibration uncertainties and the stochastic displacements from atmospheric turbulence, after correction using the techniques developed by \citet{willow}.  The result is an estimate $\sigma_{ij}$ of the RMS uncertainty along each of the $\alpha$ and $\delta$ axes for the position of tracer $i$ in exposure $j$.

The uncertainty in the impulse $I$ applied to tracer $i$ is estimated using the Fisher information matrix $\matF_i$ for its observations, defined as
\begin{equation}
  (\matF_i)_{uv} = \sum_j \frac{1}{\sigma^2_{ij}} \left[ \frac{\partial \alpha(t_j)}{\partial q_u}  \frac{\partial \alpha(t_j)}{\partial q_v} \cos^2 \delta(t)+ 
      \frac{\partial \delta(t_j)}{\partial q_u}  \frac{\partial \delta(t_j)}{\partial q_v} \right],
    \label{eq:fisher}
  \end{equation}
where $u,v$ run over the 8 elements of the parameter vector $\vecq$ for this MBA, and the sum runs over \lsst\ observations of tracer $i$.

The Cramer-Rao theorem then provides a lower bound on the uncertainty $\sigma_I$ of the inference of the impulse on a given tracer:
\begin{eqnarray}
  \sigma^2_I & =  \left[ \matF^{-1}\right]_{II} & \quad \text{marginalizing over}\, \vecx_0, \vecv_0, A_2, \nonumber\\
  \sigma^2_I & =  \left[ \matF_{A_2I}^{-1}\right]_{II} &  \quad \vecx_0, \vecv_0\, \text{known, marginalizing}\, A_2,
                      \label{eq:sigIi} \\
  \sigma^2_I & =  \left(\matF_{II}\right)^{-1} & \quad \vecx_0, \vecv_0, A_2\,\text{known.} \nonumber
\end{eqnarray}
The first case corresponds to using only \lsst\ data to estimate the impulse $I_i$ simultaneously with the initial state vector $(\vecx_0,\vecv_0)$ and Yarkovsky strength $A_2$; the third case is the limit where $\vecx_0, \vecv_0,$ and $A_2$ are known precisely from pre-\lsst\ observations;
and the middle case is when pre-\lsst\ data give $\vecx_0$ and $\vecv_0$ but not $A_2.$
The notation $\matF_{A_2I}$ indicates truncation of the full Fisher matrix to the $2\times2$ submatrix for parameters $A_2$ and $I$.  We will label these three cases as
``\lsst\ only,'' ``Yarkovsky unknown,'' and ``initial known.'' 

The $\sigma_{Ii}$ determined from simulated \lsst\ data of a tracer is independent of the amplitude of the impulse $I,$ meaning it is also independent of the mass of the deflector, the impact parameter $|\vecb|,$ and the relative velocity $v.$  This means that our 6800 tracers are effectively a random sample of the $\sigma_I$ available from all of the known MBAs.  The forecasted $\sigma_I$ will depend primarily on the apparent magnitude $m$ of the tracer, which determines the \lsst\ measurement errors; also rather heavily on the date of the impulse relative to the start and end of the survey; and on the direction of the impulse.  Atop these deterministic factors, there is substantial stochasticity in the number and quality of a tracer's observations by \lsst.

\begin{figure}
  \centering
  \includegraphics[width=\textwidth]{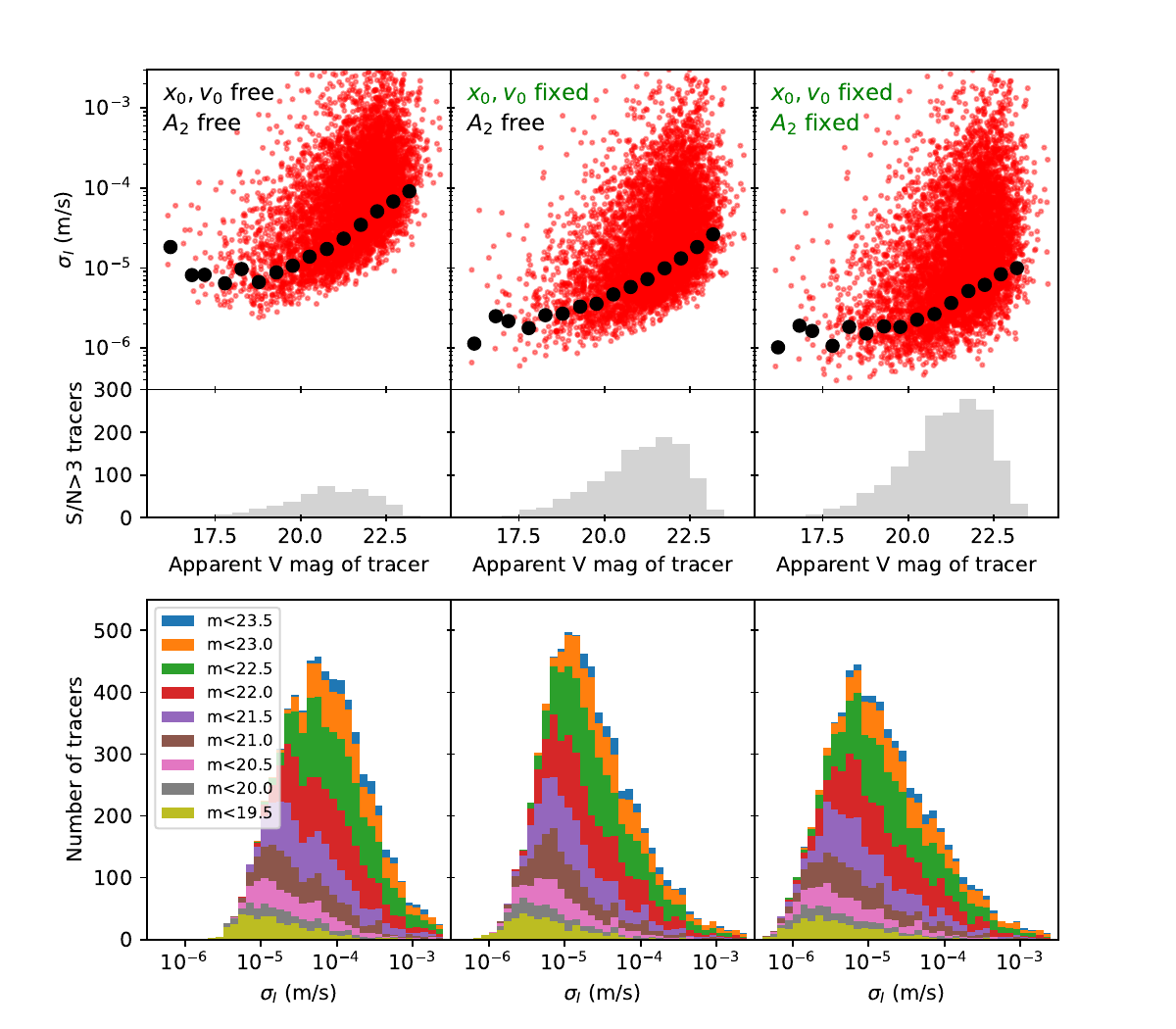}
  \caption{Each plot gives information about the uncertainty $\sigma_I$ on the applied impulse that is forecasted for observations of each tracer MBA of a mutual encounter.  The columns assume progressively more information available from pre-\lsst\ data about the tracers' initial state vectors and Yarkovsky accelerations when going from left to right, as denoted in the text atop each column.
The top row of plots is a scatter diagram (red dots) of $\sigma_I$ vs the apparent magnitude of the tracer; the black dots trace the mean vs apparent magnitude.  The middle panels give the distribution of the apparent magnitudes of those tracers which determine their deflectors' masses with $S/N>3,$ \ie\ the distribution of apparent magnitudes of the useful tracers.  The bottom panels plot the distributions of $\sigma_I$ for tracers of various magnitudes.}
  \label{fig:sigmaI}
\end{figure}

Figure~\ref{fig:sigmaI} plots several properties of the $\sigma_I$ values for the tracer population.  Of note, the information-content mean $\sigma_I,$ defined as  $\left\langle \sigma_I^{-2}\right\rangle^{-1/2},$ does degrade for fainter tracers, roughly as $\sigma_I \propto 1/\text{flux}$ (upper row of the Figure).  The middle row shows that this leads the distribution of useful tracers to peak at magnitudes 21--22, roughly where the raw MBA counts peak, suggesting that discoveries of a large number of fainter MBAs by \lsst\ could increase the number of useful encounters and improve over the results we derive.  We also see that there is substantial degeneracy between $I$ and the initial state vector in the orbit fits, because fixing the latter, as we move from left to central column, lowers the uncertainty on the former by 3--5$\times.$  Advance knowledge of the Yarkovsky strength, indicated by moving from the center to right-hand column, improves $\sigma_I$ by another factor of $\approx2,$ for the fainter tracers, but very little for the brightest ones.\footnote{It is worth noting that there are currently no claims of detections of $A_2$ for individual MBAs, but this should change when the completed \gaia\ and \lsst\ data are coupled.}

The bottom row of Figure~\ref{fig:sigmaI} also confirms that our initial choice to investigate only those mutual events causing $I>2.2\times10^{-6}\,{\rm m}\,{\rm s}^{-1}$ has not excluded any significant number of encounters that could have been useful.  In the left panel, we see that none of the tracers are capable of detecting this chosen $I_{\rm min}$ when only \lsst\ data are used.  The center and right panels shows that even with strong pre-\lsst\ orbital knowledge, only a few percent of tracers have $\sigma_I < I_{\rm min},$ and baslically none could detect $I_{\rm min}$ with $S/N>3.$  The lowest $\sigma_I$ values are, as expected, attained by brighter tracer MBAs.

\section{Resulting deflector mass uncertainties}
\label{results}
For each deflector-tracer encounter, we have derived $\sigma_I$ for the tracer's \lsst\ astrometry.  This can be propagated into an uncertainty $\sigma_M=(bv/2G)\sigma_I$ using the circumstances of the encounter.  Because the more massive deflectors are involved in multiple deflections with $I>I_{\rm min},$ we examine results for two versions of $\sigma_M$ for deflector $d:$
\begin{eqnarray}
  \sigma_M & = & {\rm min}_i\left(\sigma_{M,i}\right) \quad \text{(best tracer),} \\
  \sigma_M & = & \left(\sum_i \sigma_{M,i}^{-2}\right)^{-1/2} \quad \text{(all tracers)}.
                 \label{eq:sigM}
\end{eqnarray}
In both cases, we operate on the tracers $i$ that are deflected by the chosen MBA.  The ``best tracer'' case picks the single most informative tracer's precision, while the ``all tracers'' case combines the power of all encounters with $I>I_{\rm min}$ for up to the 20 highest-$I$ encounters. We find that combining information from the several most-informative tracers improves $\sigma_M$ by a modest factor ($\approx 2$) over using the single best tracer, and in further analysis we will combine information from up to 20 mutual events surpassing the $I_{\rm min}$ threshold.  Gains from $>20$ tracers per deflector are minimal.

\begin{figure}
  \centering
  \includegraphics[width=0.6\textwidth]{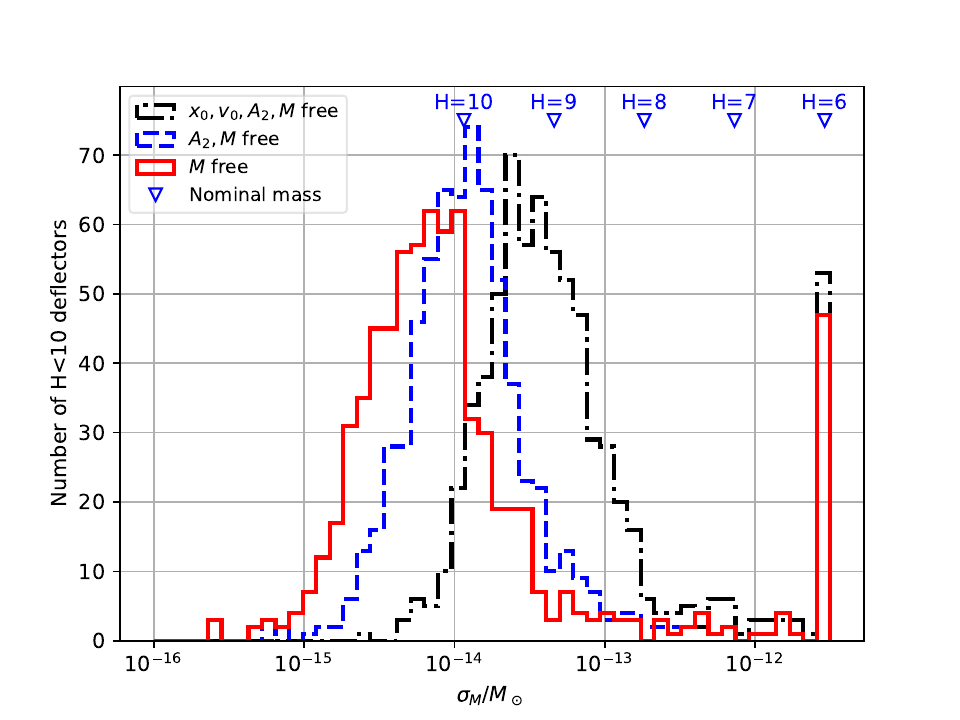}
  \caption{The distribution of forecasted $\sigma_M,$ the uncertainty on the mass of the deflector, is plotted for all of the 725 MBA deflectors with $H<10.$  The pileup at the right-hand edge represents the few objects with large uncertainties or with no encounters generating $I>I_{\rm min}$ on a tracer.  The dash-dot black histogram is the ``\lsst\ only'' case, the dashed blue histogram assumes that pre-\lsst\ data strongly constrain the initial state vector of the tracer, and the solid red histogram shows what happens if the Yarkovsky coefficient $A_2$ is also known precisely.  Along the top row are marked nominal masses of MBAs at different $H$ values.  For \lsst-only inferences, $\sigma_M$ is near the typical mass of an $H=9$ MBA, while with initial states known, $\sigma_M$ is typically at the mass of $H=10$ MBAs.}
  \label{fig:sigmaM}
\end{figure}

Figure~\ref{fig:sigmaM} plots the distributions of $\sigma_M$ for the $H<10$ asteroids, for the ``all tracers'' case.  The distribution of $\sigma_M$ should be indepedent of the deflector mass; however our imposition of a minimum on the applied impulse means that only for the $H<10$ deflectors are we locating most of the 20 highest-$I$ encounters with tracers in our analysis.  Therefore we plot the distributions of $\sigma_M$ only for this more massive subsample of deflector MBAs, as a better representation of what $\sigma_M$ is attainable.  The median deflector has $\sigma_M\approx10^{-13.4}M_\odot$ purely from \lsst\ information, which is comparable to the mass of an $H=9$ MBA.  In the ``Yarkovsky unknown'' scenario, where pre-\lsst\ data fix the initial state vector, the $\sigma_M$ distribution peaks at $\approx10^{-13.9}M_\odot,$ which is $3.3\times$ lower and close to the mass of an $H=10$ MBA.  If the Yarkovsky term can be constrained well, this can be lowered  to $\sigma_M\approx10^{-14.1}M_\odot,$ another $1.7\times$ improvement in sensitivity.

\begin{figure}
  \centering
  \includegraphics[width=0.6\textwidth]{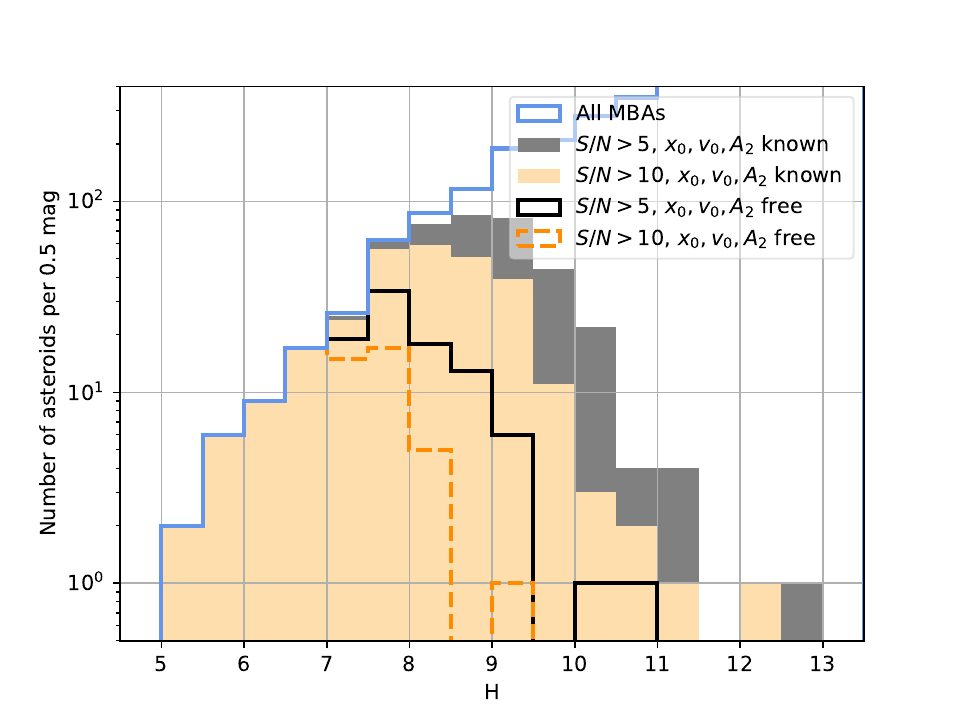}
  \caption{Histogram of the number of MBAs that are forecasted to attain $S/N>10$ or $S/N>5$ measurements of their masses by the use of \lsst\ astrometry on tracer asteroids.  The filled histograms assume that pre-\lsst\ data establish the initial state vectors and Yarkovsky strengths of the tracer MBAs.  In this case, comparing to the total differential MBA counts in blue, we find that nearly all MBAs with $H<8.5$ are measured at $S/N>5.$  If only \lsst\ observations are used, the most conservative case,  the open histograms show that $S/N>5$ is attained for nearly all $H<7.5$ MBAs.  In each case there is a tail extending 4~mag fainter in $H$ of MBAs that obtain $S/N>5$ from fortunate close encounters with a known MBA as tracer.}
  \label{fig:snhist}
\end{figure}

Figure~\ref{fig:snhist} plots our forecasts for how many MBAs will have their masses determined to $S/N>5$ or $S/N>10$ through their deflections of other MBAs.  For clarity, of our three cases ``\lsst\ only,'' ``Yarkovsky unknown,'' and ``initial known,'' we plot only the first and last.  The total number of MBAs with $S/N>5$ on their masses (assuming nominal density and albedo) in the three cases are 126, 310, and 440, respectively.  Of these, 72, 183, and 281 are forecasted to attain $S/N>10$ accuracy on their masses.  The acquisition of \lsst\ data will thus increase the number of usefully-measured MBA masses into the hundreds, from the current tens.

Regardless of the $S/N$ threshold and the extent of pre-\lsst\ data, the mass determinations follow a common pattern.  Up to some $H$ corresponding to masses several times the typical $\sigma_M,$ the mass determinations are complete.  This limiting $H$ will be between 7 and 9, depending on the $S/N$ threshold and the pre-\lsst\ observations.  But only $\approx1/2$ of the MBAs with successful mass measures are brighter than this completeness level; the other half are fainter, less massive MBAs that have fortunately close encounters with one or more known tracer MBAs, yielding $\sigma_M$ values more precise than the median.  Thus we expect to gain useful mass information on an essentially random subsample of MBAs down to $\approx100\times$ lower mass than the nearly-complete sample.

Full information on all of the forecasted encounters is given in Table~\ref{tab:events}, namely the identities of the deflector and tracer asteroids, the time and circumstances of the impulse, the nominal size of impulse given the $H$ value of the deflector, and the forecasted $\sigma_I$ of the tracer in the three scenarios of pre-\lsst\ knowledge.  Then in Table~\ref{tab:deflectors}, we combine the information from all tracers of a given deflector to give a forecast of the attainable $\sigma_M$ on each one.

The total numbers of MBAs attaining mass measurements at $S/N>1$ are 401, 683, and 777 in the three scenarios of pre-\lsst\ data.  This is a bit of an underestimate since we have not identified all mutual encounters capable of this level of precision, especially if $S/N>1$ can be attained by combining data from multiple tracers.  As explored in \citet{brownian}, these numbers are important if we aim to produce the most accurate possible ephemerides for objects in the asteroid belts, because the level of error in such ephemerides is determined by the masses of the largest asteroids that do not have any useful observational information on their mass.  For objects with $M<\sigma_M,$ the mass estimate from converting $H$ to a mass with assumed albedo and density is more accurate than the estimate from mutual events.

\begin{deluxetable}{ccccccccc}
  \tablewidth{0pt}
  \tablecaption{Mutual encounters with impulse $>2.2\times10^{-6}$~m/s and $b<0.1$~au}
  \tablehead{
   \colhead{Deflector} & \colhead{Tracer} & \colhead{MJD} & \colhead{Impact vector $b$} &
    \colhead{Relative $v$} & \colhead{Impulse} & \multicolumn{3}{c}{$\sigma_I$\tablenotemark{a}}  \\
    \colhead{(MPC number)} &\colhead{(MPC number)} & & ($10^{-3}\,\AU$, ICRS) & (km/s) &  (m/s) & \multicolumn{3}{c}{($10^{-6}$~m/s)} } 
  \startdata
  00003  & b5337  & 62737.1 & (+6.7931,+5.6376,-4.1305) & 1.59 & $1.0\times10^{-3}$ & 14.1 & 5.1 & 4.0 \\
00003  & K16Q96 & 63526.6 & (-2.5392,-2.5206,-2.9117) & 3.27 & $1.1\times10^{-3}$ & 41.9 & 29.7 & 11.8 \\
00003  & o2558  & 63497.3 & (+0.5945,+2.9759,-0.3743) & 4.91 & $1.1\times10^{-3}$ & 116.8 & 108.2 & 106.6 \\
00003  & H6785  & 63631.5 & (-0.6189,-4.5772,-1.7071) & 3.02 & $1.1\times10^{-3}$ & 285.3 & 268.7 & 264.2 \\
00003  & K11BD0 & 62485.7 & (+0.2078,+5.0813,-3.1908) & 2.43 & $1.1\times10^{-3}$ & 77.1 & 19.6 & 16.1 \\
\multicolumn{9}{c}{\nodata} \\
17822  & \~081n  & 63907.1 & (-0.0689,-0.0165,-0.1449) & 0.23 & $3.1\times10^{-6}$ & 102.7 & 70.3 & 65.0 \\
26768  & y0276  & 62128.8 & (-0.0069,+0.0249,-0.0099) & 0.91 & $2.6\times10^{-6}$ & 61.6 & 21.0 & 7.9 \\
55167  & t2034  & 62543.7 & (+0.0158,-0.0141,-0.0027) & 0.51 & $4.3\times10^{-6}$ & 40.6 & 24.0 & 23.4 \\
24680  & K20M53 & 62203.2 & (+0.0000,+0.0017,-0.0005) & 3.14 & $2.7\times10^{-6}$ & 326.4 & 99.1 & 73.1 \\
 \enddata
  \tablenotetext{a}{The three columns give values in scenarios where only \lsst\ data are used, where pre-\lsst\ data determine the tracer asteroids' initial state vectors, and where the Yarkovsky acceleration $A_2$ is also known, respectively.}
\tablecomments{Table \ref{tab:events} is published in its entirety in the machine-readable format.
  A portion is shown here for guidance regarding its form and content.}
\label{tab:events}
\end{deluxetable}

\begin{deluxetable}{ccccccccc}
  \tablewidth{0pt}
  \tablecaption{Forecasted mass uncertainties for deflector MBAs with $S/N>5$}
  \tablehead{
    \colhead{MPC Number} & \colhead{Designation} & \colhead{$H$} & \multicolumn{3}{c}{$\sigma_M/M_\odot$\tablenotemark{a}} & \multicolumn{3}{c}{Nominal $S/N$\tablenotemark{a}}  }
  \startdata
3 & Juno & 5.18 & $6.0\times10^{-14}$ & $3.2\times10^{-14}$ & $1.1\times10^{-14}$ & 149.4 & 285.2 & 834.0 \\
5 & Astraea & 6.99 & $8.1\times10^{-15}$ & $4.9\times10^{-15}$ & $3.8\times10^{-15}$ & 91.8 & 150.6 & 195.5 \\
6 & Hebe & 5.61 & $6.2\times10^{-14}$ & $4.2\times10^{-14}$ & $1.5\times10^{-14}$ & 80.7 & 118.8 & 322.2 \\
7 & Iris & 5.67 & $2.5\times10^{-14}$ & $6.1\times10^{-15}$ & $2.3\times10^{-15}$ & 186.0 & 747.3 & 2009.1 \\
8 & Flora & 6.62 & $6.3\times10^{-15}$ & $3.9\times10^{-15}$ & $1.1\times10^{-15}$ & 195.1 & 316.8 & 1088.6 \\
\multicolumn{9}{c}{\nodata} \\
1282 & Utopia & 10.28 & $7.1\times10^{-14}$ & $4.0\times10^{-15}$ & $1.5\times10^{-15}$ & 0.1 & 2.0 & 5.1 \\
1321 & Majuba & 10.14 & $4.5\times10^{-15}$ & $3.0\times10^{-15}$ & $1.5\times10^{-15}$ & 2.1 & 3.2 & 6.6 \\
1628 & Strobel & 10.23 & $1.6\times10^{-15}$ & $9.3\times10^{-16}$ & $6.3\times10^{-16}$ & 5.2 & 9.0 & 13.5 \\
2535 & Hameenlinna & 12.49 & $9.8\times10^{-16}$ & $1.4\times10^{-16}$ & $3.7\times10^{-17}$ & 0.4 & 2.7 & 10.1 \\
  \enddata
  \tablenotetext{a}{The three columns give values in scenarios where only \lsst\ data are used, where pre-\lsst\ data determine the tracer asteroids' initial state vectors, and where the Yarkovsky acceleration $A_2$ is also known, respectively.}
\tablecomments{Table \ref{tab:deflectors} is published in its entirety in the machine-readable format.
  A portion is shown here for guidance regarding its form and content.}
\label{tab:deflectors}\end{deluxetable}

\section{Discussion}
The conclusions of our forecast of \lsst\ constraints on mutual encounters of asteroids can be summarized in a few different domains.  In terms of ability to measure an impulse $I$ occurring sometime during \lsst\ years 2--9, the typical information from observations of an MBA with apparent $V$-band magnitude $m<19.5$ measures the impulse to $\sigma_I\approx7\times10^{-6}\,{\rm m}\,{\rm s}^{-1}$ if we use only data from \lsst\ to solve simultaneously for the initial state vector of the tracer, its non-gravitational acceleration $A_2,$ and the applied impulse $I$.  If pre-\lsst\ observations constrain the initial state strongly, this improves to $\approx2\times10^{-6}\,{\rm m}\,{\rm s}^{-1}$, and 10\% or so lower if the Yarkovsky uncertainty is removed.
These numbers degrade for fainter tracer MBAs, which have higher observational astrometric uncertainties and larger Yarkovsky accelerations, becoming $\approx10\times$ worse at $m=23$~mag.

In terms of the uncertainty $\sigma_M$ on the mass of a deflector that is attained by
\lsst\ astrometry of the $\sim10$ tracer MBAs on which it imparts the largest impulses,
we find a median value $\sigma_M=4\times10^{-14} M_\odot$ using only \lsst\ data, improving to $1.3\times10^{-14}M_\odot$ and $7\times10^{-15} M_\odot$ as pre-\lsst\ data become capable of constraining the initial state vector and then the Yarkovsky $A_2.$
The value of $\sigma_M$ for a given deflector varies, however, by more than a factor 10,  depending upon the ``luck'' of a given deflector MBA in having encounters at small distances $b$ and relative velocities $v$ with brighter tracers.

In terms of the number of MBAs with masses measured to $S/N>5,$ we forecast 126, 310, and 440 in the three scenarios of pre-\lsst\ data quality, with $>1/2$ of these attaining $S/N>10.$ 
The largest 10--50 bodies (beyond Vesta, Ceres, and Pallas) will have their masses measured to 1\% accuracy or better.
Typically the $S/N>5$ mass determinations will be nearly complete for MBAs down to 5 times the median $\sigma_M,$ and this ``complete'' set will comprise about half of the determinations.  A random selection of lower-mass objects will obtain favorable encounter circumstances and acquire useful mass determinations.
We have provided tables of the deflectors with prospects for mass detection, and the nature of the encounters that inform their masses: the tracer involved, geometry of the event, and forecasted accuracies on the impulse applied to each tracer.

Our forecasts incorporate a simplified model of the Yarkovsky effect as an azimuthal acceleration given by a time-independent $A_2$ value for each tracer MBA.  In practice it may be necessary to incorporate more complex radiation forces, but we believe the single parameter captures most of the resultant uncertainties in deflector mass.  Another effect that could interfere with attaining the Fisher-forecasted uncertainties on deflector masses is that each tracer is subject to deflections by all of the millions of other MBAs, which are individually undetectable in its trajectory, but combine to cause a Brownian motion that would be a source of astrometric noise.  In \citet{brownian}, we estimate this noise and find it would indeed be negligible for present purposes.  This result is perhaps expected, because in the current work we showed that only $\approx0.5\%$ of MBAs are subjected to a detectable impulse during a decade's observations.

The forecasts show that there is potentially much to gain by incorporation of pre-\lsst\ astrometry of the tracer MBAs into the inferences on the impulses that occur during \lsst.  Such data would need to be placed onto the \gaia\ reference frame and have reliable uncertainties estimated, if the original reported positions did not already include these.  We have not delved into this exercise but it would be essential to getting the most out of \lsst.  Of the 440 deflectors with the potential for reaching $S/N>5$ mass detections in the best-case pre-\lsst\ scenario, 311 of them deflect at least one tracer that has measurements in the \gaia\ Focused Product Release table of solar system object positions \citep{gaiafprsso}.  Most of the known MBAs have been measured multiple times in well-characterized ground-based surveys in 2010--2025, so the prospects for attaining the forcasts in the ``Yarkovsky unknown'' scenario are good.

Our forecasts are pessimistic in the sense that they do not account for information that could be acquired from tracer MBAs that \lsst\ will discover.  Since there are, by definition, no pre-\lsst\ MPC entries for these sources, any pre-\lsst\ data would have to be ``precovered'' from re-examination of earlier catalogs and images, and will necessarily be lower quality, if they can be found at all.

The computational burden of determining the 100's of MBA masses that are detectable in \lsst\ data should be small.  We have seen that only a few thousand out of the many millions of \lsst-detectable asteroids will need to have precision orbit fitting executed.   \citet{brownian} shows that it is only necessary to include the masses of a few hundred MBAs as free parameters in a global ephemeris fit in order to reduce errors from unmodelled masses to insignificant levels.  This will be a relatively easy task.

Finally, we note that it should be possible to estimate the \emph{mean} relation of mass to $H$ for classes of MBAs divided into orbital and/or photometric classes, even when \emph{individual} objects' masses have insufficient $S/N$ for detection.  This could be done by summing (``stacking'') the information from all of their collective gravitational encounters with other MBAs.  This would allow characterization of the surface/composition properties across color and dynamical space for smaller bodies than those cataloged in this work.

In summary we find that \lsst\ should boost the number of MBAs with usefully measured masses ($S/N\gtrsim5$) from the 10's into the 100's, including nearly complete coverage of bodies with $H\lesssim 9.$  With such studies, time is on the observer's side: the number of measurable encounters increases with time, the size of the perturbations from a given impulse grows with time, the ability to distinguish an impulse from the Yarkovsky effect improves with time, and the continued observations improve the sensitivity of the data.

\begin{acknowledgments}
  This work was supported by NSF grant AST-2205808.  We thank Zachary Murray for very helpful discussions. We also acknowledge the valuable work of the Rubin Observatory Survey Cadence Optimization Committee in creating and publishing simulations of the \lsst.  This research has made use of data and/or services provided by the International Astronomical Union's Minor Planet Center. 
\end{acknowledgments}

\newpage
\bibliographystyle{aasjournal}
\bibliography{references}

\end{document}